\documentclass[prl,twocolumn,aps,superscriptaddress,preprintnumbers,nofootinbib]{revtex4}
\usepackage{graphicx}
\usepackage{amsfonts}
\usepackage{amssymb}
\usepackage{amsmath}
\usepackage{hyperref}
\usepackage{slashed}
\usepackage{multirow}
\usepackage{fancyhdr}
\usepackage{appendix}
\usepackage{subfigure}
\usepackage{times}

\def\y{\psi}

\DeclareMathOperator{\hz}{h\relax{\kern-.15em}z}
\DeclareMathOperator{\pz}{\psi\relax{\kern-.15em}z}

\usepackage[english]{babel}
\usepackage{amssymb}
\usepackage{amsfonts}
\usepackage{amsmath}
\usepackage{graphicx}
\usepackage{epsfig}
\usepackage{psfrag}
\newcommand{\be}{\begin{equation}} \newcommand{\ee}{\end{equation}}
\newcommand{\bea}{\begin{eqnarray}} \newcommand{\eea}{\end{eqnarray}}
\newcommand{\beann}{\begin{eqnarray*}}  \newcommand{\eeann}{\end{eqnarray*}}
\newcommand{\bfig}{\begin{figure}} \newcommand{\efig}{\end{figure}}
\newcommand{\ba}{\begin{array}} \newcommand{\ea}{\end{array}}
\newcommand{\bcen}{\begin{center}} \newcommand{\ecen}{\end{center}}
\newcommand{\btab}{\begin{tabular}} \newcommand{\etab}{\end{tabular}}

\newcommand{\matt}{\left ( \begin{array}{ccc}}
    \newcommand{\ematt}{\end{array} \right )} \newcommand{\matf}{\left ( \begin{array}{cccc}}
    \newcommand{\ematf}{\end{array} \right )} \newcommand{\vect}{\left ( \begin{array}{c}}
    \newcommand{\evect}{\end{array} \right )}    \def\beqn{\begin{eqnarray}}
 \def\eeqn{\end{eqnarray}}  
     
\newtheorem{Proposition}{Proposition}[section]

\newtheorem{Theorem}{Theorem}[section]
\newtheorem{Lemma}{Lemma}[section]
\newtheorem{Corrolary}{Corrolary}[section]

\newcommand{\bp}{\begin{Proposition}}   \newcommand{\ep}{\end{Proposition}}
\newcommand{\bt}{\begin{Theorem}}   \newcommand{\et}{\end{Theorem}}
\newcommand{\bl}{\begin{Lemma}}     \newcommand{\el}{\end{Lemma}}
\newcommand{\bc}{\begin{Corrolary}} \newcommand{\ec}{\end{Corrolary}}

\begin{document}

\preprint{IFT-UAM/CSIC-13-095}

\title{Holographic s+p Superconductors}

\author{Irene Amado}\email{irene.r.amado@gmail.com}
\affiliation{Department of Physics, Technion, Haifa 32000, Israel}

\author{Daniel Are\'an}\email{darean@ictp.it}
\affiliation{International Centre for Theoretical Physics (ICTP), Strada Costiera 11, I 34014 Trieste, Italy}
\affiliation{INFN, Sezione di Trieste, Strada Costiera 11, I 34014 Trieste, Italy}

\author{Amadeo Jimenez-Alba}\email{amadeo.j@gmail.com}
\affiliation{Instituto de F\'\i sica Te\'orica IFT-UAM/CSIC, Universidad
Aut\'onoma de Madrid,
28049 Cantoblanco, Spain}

\author{Luis Melgar}\email{luis.melgar@csic.es}
\affiliation{Instituto de F\'\i sica Te\'orica IFT-UAM/CSIC, Universidad
Aut\'onoma de Madrid,
28049 Cantoblanco, Spain}

\author{Ignacio Salazar Landea}\email{peznacho@gmail.com}
\affiliation{ Instituto de F\'\i sica La Plata (IFLP) and Departamento de F\'\i sica Universidad Nacional de La Plata, CC 67,
1900 La Plata, Argentina}
\affiliation{International Centre for Theoretical Physics (ICTP), Strada Costiera 11, I 34014 Trieste, Italy}

\begin{abstract}
We study the phase diagram of a holographic model realizing a $U(2)$ global symmetry on the boundary and show that at low temperature a phase with both scalar $s$ and vector $p$ condensates exists. This is the $s$+$p$-wave phase where the global $U(2)$ symmetry and also the spatial rotational symmetry are spontaneously broken. By studying the free energy we show that this phase is preferred when it exists.
We also consider unbalanced configurations where a second chemical potential is turned on.
They present a rich phase diagram
characterized by the competition and coexistence of the $s$ and $p$ order parameters.
\end{abstract}

\pacs{ }

\maketitle


\section{ Introduction}
An interesting problem in the arena of unconventional superfluids and superconductors is that of the competition
and coexistence of different order parameters \cite{zhang}. A paradigmatic example in the realm of superfluids is
that of ${}^3$He. At low temperature ${}^3$He  presents two distinct superfluid phases, denoted as $A$ and
$B$ phases \cite{he3phases}. ${}^3$He-$B$ is the low temperature (and low pressure) phase and it
corresponds to a $p$-wave superfluid, where the order parameter transforms as a vector under spatial rotations.
${}^3$He-$A$ is the higher temperature (and pressure) superfluid phase. It is a chiral $p$-wave superfluid
whose order parameter is a complex vector, and time-reversal and parity symmetry are spontaneously broken.
In the domain of unconventional superconductors it has been shown in \cite{sip} that for doped three dimensional
narrow gap semiconductors such as ${\rm Cu}_x {\rm Bi}_2{\rm Se_3}$ and
${\rm Sn}_{1-x}{\rm In}_x{\rm Te}$ there is a competition between $s$ and $p$-wave superconducting states.
Dialing the coupling constants of the two different channels (corresponding to the $s$ and $p$ pairings) leads to
a phase diagram where both a $p$ and an $s$-wave phase exist. Moreover, at the interface of both phases a new
$p$+$is$ state appears. The order parameter for this phase is the combination of a vector and a pseudoscalar, and
breaks both time-reversal and parity symmetry, making this state an interesting example of a topological
superconductor\footnote{This is actually an example of an axionic state of matter. This $p$+$is$ phase belongs
to the class D in the classification \cite{ludwig} of 3D topological superconductors. It possesses gapped Majorana
fermions as edge states which give rise to an anomalous surface thermal Hall effect. It would be very interesting
to realize holographically this axionic superconducting state (see \cite{Zayas:2011dw} for a holographic
time-reversal symmetry breaking $p$+$ip$ superconductor).}.

The AdS/CFT correspondence has succeeded in constructing a holographic version of superconductivity \cite{Gubser:2008zu,Hartnoll:2008vx} (for comprehensive reviews see \cite{Hartnoll:2009sz,Horowitz:2010gk}).
Furthermore, holographic models of $s$ \cite{Hartnoll:2008kx}, $p$  \cite{phsc} and $d$-wave
\cite{dhsc} superconductors; which have scalar, vector, and spin-2 order parameters respectively, have been developed in the last years.
Coexistence and competition of several order parameters has also been addressed holographically in
\cite{Basu:2010fa, Zayas:2011dw, Musso:2013ija,Cai:2013wma, Wen:2013ufa, Nie:2013sda,Amoretti:2013oia,Donos:2013woa}\footnote{In
\cite{Nie:2013sda}, which appeared when this work was being completed, a holographic $s$+$p$-wave phase was also found.}.

In this letter, building upon a model constructed in \cite{we}, we develop a holographic dual of a superconductor
with both $s$-wave and $p$-wave condensates.
Subsequently we study the phase diagram of unbalanced mixtures (where two chemical potentials are turned on)
finding a competition of $s$, $p$, and $s$+$p$-wave superconducting phases.

In \cite{we} a holographic dual of a two-component superfluid \cite{halperin} was constructed, consisting on a scalar doublet
charged under a $U(2)$ gauge field living in a planar Schwarzschild Black Hole (BH) geometry. Switching on a
chemical potential along the overall $U(1)\subset U(2)$, the system becomes unstable towards the condensation
of the scalar doublet. The appearance of the scalar condensate spontaneously breaks the $U(2)$ symmetry down
to $U(1)$, signaling a phase transition to an $s$-wave superfluid phase. In this phase two different charge
densities are present in the system, corresponding to the two $U(1)$s inside the $U(2)$, hence realizing a
holographic two-component superfluid. It was also found that the $s$-wave superfluid phase is unstable
at low temperatures and argued that this instability signaled the appearance of a non-trivial $p$-wave order
parameter. In the present paper we confirm that prediction and explicitly construct the solutions in which
condensation of a vector mode breaks the remaining $U(1)$ and gives rise to a new phase with two condensates:
the $s$+$p$-wave holographic superconductor. The study of these new solutions allows us to determine the phase
diagram of the two-component superfluid.

If one works in the grand canonical ensemble, where the chemical potential of the boundary theory is held fixed,
the temperature of the system is given by $T \propto 1/\mu$, where $\mu$ is a dimensionless chemical potential
related to that of the boundary theory by rescalings. The final picture is the following: at small enough chemical
potential $\mu$ (high temperature) the system is in the normal phase where no condensate is present.
For $\mu$ greater than a critical value $\mu_s$ the scalar field acquires an expectation value and the system
enters the $s$-wave superfluid phase. Going to even larger chemical potential a new phase transition happens:
at $\mu_{sp}>\mu_s$ a vector condensate appears and for $\mu>\mu_{sp}$ the system is in an $s$+$p$-wave
phase with both scalar and vector non-vanishing order parameters.

Finally, we shall study new configurations of the system where the two chemical potentials corresponding
to the two $U(1)$s $\subset U(2)$ are switched on.
This setup, where the $U(2)$ is explicitly broken to $U(1)\times U(1)$, realizes an unbalanced mixture,
characterized by the presence of two species of charges with different chemical potentials.
Examples of such systems are unbalanced Fermi mixtures \cite{imfermi}, and QCD at finite baryon and isospin
chemical potential \cite{imqcd}. Moreover, unbalanced superconductors are interesting systems where anisotropic
and inhomogeneous phases are expected to appear \cite{loffrv,loff}. Holographic realizations of unbalanced
systems where only one kind of order parameter can be realized have been constructed in
\cite{Bigazzi:2011ak, Erdmenger:2011hp}. Here we construct new solutions of the system in \cite{we}
corresponding to unbalanced mixtures that allow for competition of different order parameters.
We determine its phase diagram as a function of the two chemical potentials and find that $s$-wave,
$p$-wave and $s$+$p$-wave phases exist.

\section{ The holographic two-component superfluid}

Let us consider the holographic model of a multi-component superfluid consisting of a scalar doublet charged
under a $U(2)$ gauge field living in a $3+1$ dimensional Schwarzschild-AdS black brane geometry constructed
in \cite{we}\footnote{A similar model was introduced in \cite{Krikun:2012yj} in order to describe holographic
multiband superconductors.}. The action for such a system reads
\be
S=\int d^4 x \sqrt{-g}\left(-\frac{1}{4}F^{\mu\nu}_{c}F_{\mu\nu}^c-m^2\Psi^\dagger \Psi-(D^\mu\Psi)^\dagger D_\mu\Psi\right)\,,
\label{accion}
\ee
with
\bea
&&\Psi=\sqrt{2} \begin{pmatrix}
\lambda \\
\psi
\end{pmatrix}\;,\quad
D_{\mu}=\partial_{\mu}-iA_{\mu}\;,\quad
A_{\mu}=A_{\mu}^{c}T_c\,, \\
&&
 T_0=\frac{1}{2}\mathbb{I}\;,\quad
 T_i=\frac{1}{2}\sigma_i\,.
\eea
The system lives in the Schwarzschild-AdS background
\bea
&&ds^2=-f(r)dt^2 +\frac{dr^2}{f(r)} + r^2 \left(dx^2+dy^2\right)\,, \nonumber  \\
&& f(r)=r^2\left(1-\frac1{r^{3}}\right) \,,
\label{metric}
\eea
where we have set the radius of AdS and of the horizon to $L=r_h=1$, by using the scaling symmetries of the system. We work in the decoupling limit, in which the backreaction of the matter fields on the metric is negligible.

We consider the following (consistent) ansatz for the fields in our setup \cite{we}
\be
A^{(0)}_0= \Phi(r)\,,\quad   A^{(3)}_0 = \Theta(r)\,,\quad A^{(1)}_1 = w(r)\,,\quad \psi=\psi(r)\,,
\ee
with all functions being real-valued. All other fields in (\ref{accion}) are set to zero, in particular we set $\lambda=0$ without loss of generality. The resulting equations of motion read
\bea
&&\hspace{-0.5cm}\y''+\left(\frac{f'}{f}+\frac{2}{r}\right)\y'+\left(\frac{(\Phi-\Theta)^2}{4f^2}-\frac{m^2}{f}-\frac{w^2}{4r^2f}\right)\y=0\,,\label{scalar}
\nonumber\\ \\
&&\label{gauge1} \hspace{-0.5cm}\Phi''+\frac{2}{r} \Phi'-\frac{\y^2}{f}(\Phi-\Theta)=0\,,\\
&&\label{gauge2}\hspace{-0.5cm} \Theta''+\frac{2}{r}\Theta'+\frac{\y^2}{f}(\Phi-\Theta)-\frac{w^2}{r^2f}\Theta=0\,,\\
&&\label{gauge3}\hspace{-0.5cm} w''+\frac{f'}{f}w'+\frac{\Theta^2}{f^2}w-\frac{\y^2}{f}w=0\,.
\eea
In what follows we choose the scalar to have $m^2 = -2$ and the corresponding dual operator to have mass dimension 2.

The UV asymptotic behavior of the fields, corresponding to the solution of equations (\ref{scalar} - \ref{gauge3}) in the limit $r\to\infty$, is given by
\bea
&&\Phi=\mu-\rho/r+O(r^{-2})\,,\\
&&\Theta=\mu_3-\rho_3/r+O(r^{-2})\,,\\
&&w = w^{(0)}+w^{(1)}/r+O(r^{-2})\,,\\
&&\psi = \psi^{(1)}/r+\psi^{(2)}/r^2+O(r^{-3})\,,
\eea
where, on the dual side, $\mu$ and $\rho$ are respectively the chemical potential and charge density corresponding
to the overall $U(1)\subset U(2)$ generated by $T_0$, whereas $\mu_3$ and $\rho_3$ are the chemical potential and
charge density corresponding to the $U(1)\subset SU(2)$ generated by $T_3$. $\psi^{(1)}$ is the source of a scalar operator of dimension 2,
while $\psi^{(2)}$ is its expectation value. Finally $w^{(0)}$ and $w^{(1)}$ are the source and vev of the current operator
$J_x^{(1)}$ (recall that $A^{(1)}_\mu$ is dual to the current $J_\mu^{(1)}$).  Notice that in a background where $w(r)$
condenses the $SU(2)\subset U(2)$ is spontaneously broken, and moreover spatial rotational symmetry is spontaneously broken
too.

\section{The s+p-wave holographic superconductor}
We are looking for solutions of the equations (\ref{scalar} - \ref{gauge3}) where $\psi$, $w$, or both acquire non-trivial profiles. We want them to realize spontaneous symmetry breaking so we impose that their leading UV contributions (dual to the sources of the corresponding operators) vanish. We will switch on a chemical potential $\mu$ along the overall $U(1)$, while requiring that the other chemical potential $\mu_3$ remains null. Therefore our UV boundary conditions are
\be
\psi^{(1)}=0\,,\quad w^{(0)}=0\,,\quad \mu_3 =0\,.
\label{uvconds}
\ee
In the IR regularity requires $A_t$ to vanish at the BH horizon.

Notice that after using the scaling symmetries of the system to fix the black hole parameters in (\ref{metric}), the only scale in the problem is given by the chemical potential $\mu$. In the grand canonical ensemble, in which the physical chemical potential is held fixed, the temperature is proportional to the rescaled chemical potential as $T\propto 1/\mu$. Therefore, varying $\mu$ is equivalent to changing the temperature of the system. For that reason, the results in this letter are presented in terms of $\mu$.

We have looked for numerical solutions with non-zero $\psi$ and $w$, shooting from the IR towards the UV
where we impose the boundary conditions (\ref{uvconds}). We have found the following solutions:
\newline{\bf Normal phase}: for all values of $\mu$ there exists an analytic solution where $\psi=w=\Theta=0$ and $\Phi=\mu(1-1/r)$. This solution describes the normal state of the system.
\newline{\bf $s$-wave phase}: for $\mu\geq \mu_s \approx 8.127$ we find solutions with non-zero $\psi$.
As seen in \cite{we} for these solutions the equations decouple into two sectors: one corresponding to the
Abelian holographic superconductor \cite{Hartnoll:2008vx} and the other to the unbroken $U(1)$ symmetry.
Although $\mu_3$ is zero as required in (\ref{uvconds}), both charge densities $\rho$ and $\rho_3$ are
non-vanishing and therefore a two-component s-wave superfluid is realized.
Indeed as one can see in eq. (\ref{gauge2}) a non-trivial scalar $\psi$ acts a a source for the field $\Theta(r)$,
and therefore the only pure s-wave solutions satisfying the boundary conditions (\ref{uvconds}) are those with
$\rho_3\neq0$. Hence two different charge densities ($\rho$ and $\rho_3$) corresponding to the two different
$U(1)$s $\subset U(2)$ are turned on for these solutions and it is in this sense that this phase was denoted a
two-component holographic superfluid in \cite{we}.\footnote{From eqs. (\ref{scalar} - \ref{gauge2}), one can
see that the scalar condensate is only charged under a linear combination of $\Phi$ and $\Theta$, whereas in the
absence of a vector condensate, the orthogonal combination completely decouples corresponding to the unbroken $U(1)$
gauge field. }
%
%
\newline{\bf $s$+$p$-wave phase}: for $\mu\geq \mu_{sp}\approx 20.56$ there are solutions satisfying (\ref{uvconds}) with non-zero $\psi$ and $w$. In these solutions the $U(2)$ symmetry is completely broken, and moreover since $w^{(1)}\sim \langle J_x^{(1)}\rangle$ spatial rotational symmetry is broken too. Again $\mu_3=0$ while $\rho$ and $\rho_3$ are non-vanishing, thus realizing an $s$+$p$-wave phase of a two-component superfluid.
Usually $p$-wave superconductivity is triggered by a $\mu_3$ chemical potential \cite{phsc}. Here instead the $p$
component of the $s$+$p$ superfluid is supported by the spontaneously induced charge density $\rho_3$.
For that reason no solutions with only $p$ condensate are present in this system.\footnote{It is clear from eq.
(\ref{gauge3}) that the p-wave condensate only couples directly to the $U(1)\subset SU(2)$, i.e to $\Theta(r)$.
Actually, this equation reduces to that of the standard p-wave holographic superconductor \cite{phsc} when the
scalar is switched off. As in \cite{phsc}, only a non-zero $\Theta$ in the bulk can source the vector condensate
since the coupling to the scalar $\psi$ increases the effective mass of $w$ and therefore hinders condensation.
In contrast to the standard p-wave scenario we are fixing $\mu_3=0$, but solutions with non-zero $\Theta$ are still
possible in presence of the s-wave condensate (realized by a non-zero $\psi$) as explained above.}

\begin{figure}[htp!]
\begin{center}
\includegraphics[width=3.41in]{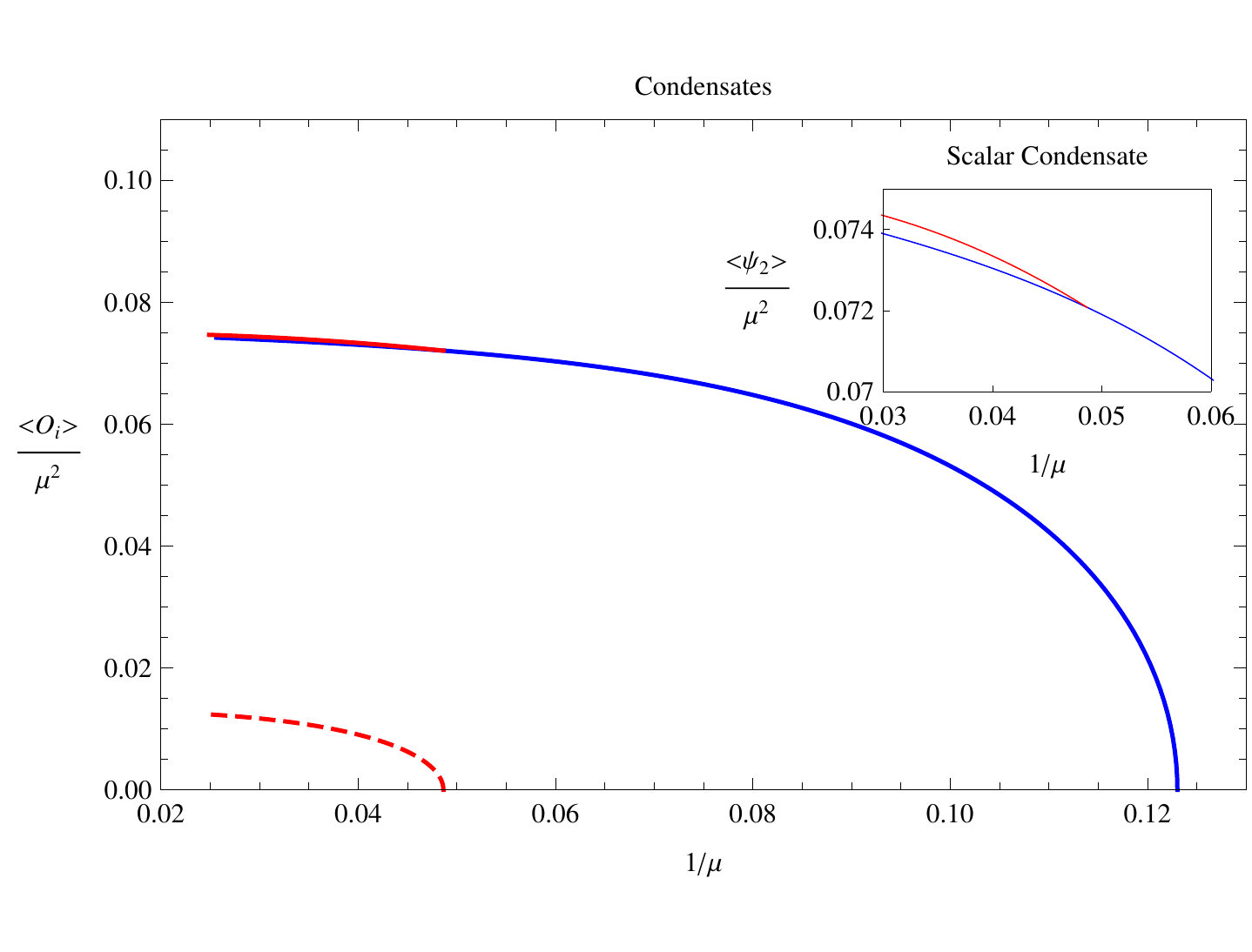}
\caption{\label{condensate} Condensates $ \psi^{(2)}$ (solid) and $w^{(1)}$ (dashed) as a function
of $1/\mu$ in the $s$-wave (blue) and $s$+$p$-wave (red) phases. The $p$ condensate appears at
$\mu_{sp}$ such that $\mu_s/\mu_{sp}=0.395$ as found in \cite{we}. The inset zooms in on the plot of
$ \psi^{(2)}$ to show the difference in the scalar condensate between the $s$ (blue) and the $s$+$p$ (red)
solutions.}
\end{center}
\end{figure}

In figure \ref{condensate} we plot the condensates $\langle O_2 \rangle \sim \psi^{(2)}$ and $\langle J_x^{(1)}\rangle\sim w^{(1)}$ as a function of the chemical potential. Notice that the solution where both condensates coexist extends down to as low $1/\mu$ (or equivalently low temperatures) as where we can trust the decoupling limit and thus neglect backreaction.

\begin{figure}[htp!]
\begin{center}
\includegraphics[width=3.41in]{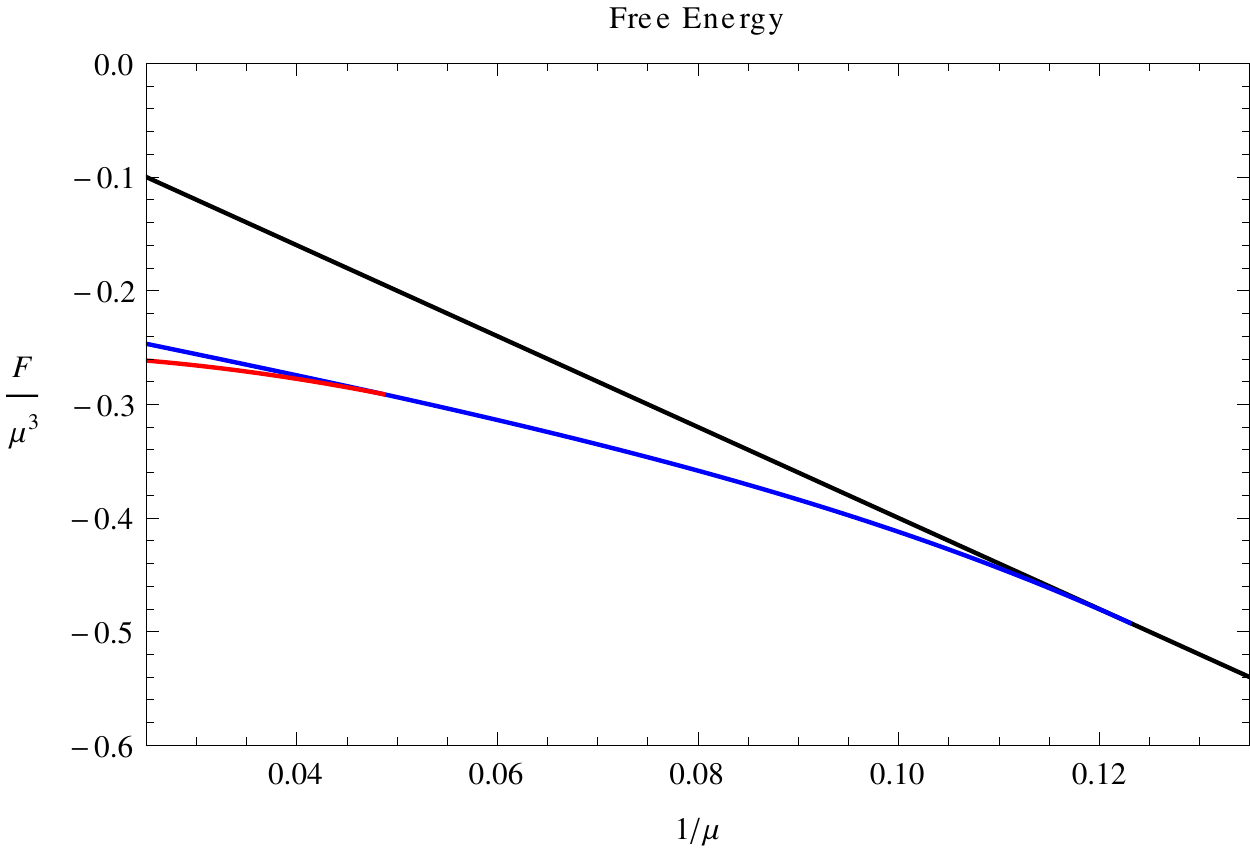}
\caption{\label{freeenplot} Free energy of the different solutions versus $1/\mu$: normal phase in black, $s$-wave phase in blue, and $s$+$p$-wave phase in red.}
\end{center}
\end{figure}

To determine the phase diagram of our system we compute the free energy of the different solutions and establish which is preferred when more than one exist. The free energy density is given by the on-shell action, and for our ansatz it reads
\bea
&&\label{freeen}F=-\frac{T}{V}S_E=-\frac12\left(\mu\,\rho+\mu_3\,\rho_3\right)\,+\\
&&\;\;+\int\frac{dr}{2f}\,(-f\,w^2\,\psi^2 +\,r^2\left(\Phi-\Theta\right)^2\,\psi^2+\frac{f}{r^2} w^2\,\Theta^2\,)\,.\nonumber
\eea
The free energy for the different solutions is shown in figure \ref{freeenplot}. At small chemical potential only the normal phase solution exists. At $\mu=\mu_s \approx 8.127$ there is a second order phase transition to the $s$-wave solution. If one keeps increasing $\mu$, at $\mu_{sp}\approx 20.56$ there is a second order phase transition from the $s$-wave phase to the $s$+$p$-wave phase. The system stays in the $s$+$p$-wave phase for $\mu>\mu_{sp}$.

\section{ Unbalanced Superconductors }
In this section we relax the condition $\mu_3 =0$ and study the phase diagram of the system as a function of
$\mu$ and $\mu_3/\mu$. Notice that turning on a second chemical pontential means to explicitly break
$U(2)\rightarrow U(1)\times U(1)$. The system can now be interpreted as a holographic dual to an unbalanced
mixture \cite{Erdmenger:2011hp,Bigazzi:2011ak}.

Now that the $U(2)$ is explicitly broken, we can not generically impose that $\lambda=0$ by using gauge
transformations.  Therefore, in principle both components of the scalar doublet may condense.
In \cite{Krikun:2012yj} it was studied which option is thermodynamically favored. Following their analysis,
choosing the condensate to be on the lower component forces us to set $\mu_3/\mu<0$ for the solutions to
be stable.

\begin{figure}[htp]
\begin{center}
\includegraphics[width=3.4in]{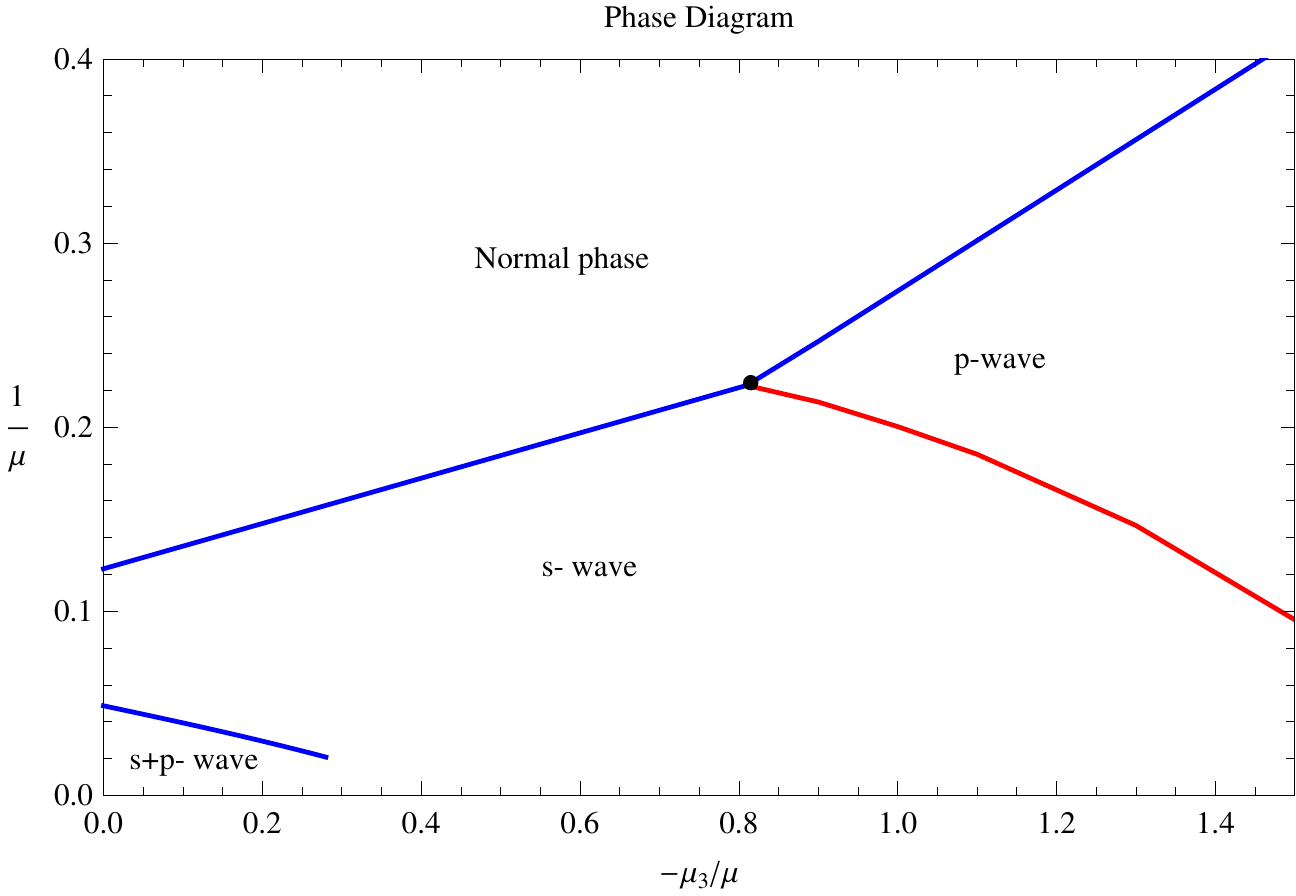}
\caption{\label{phasediagram} Phase diagram of the unbalanced system as a function of $1/\mu$ and
$\mu_3/\mu$. Second order phase transitions are denoted by blue lines, whereas the red line corresponds to a
first order phase transition.}
\end{center}
\end{figure}

The UV boundary conditions now read
\be
\psi^{(1)}=0\,,\quad w^{(0)}=0\,.\label{uvconds2}
\ee
As before we use numerical integration to solve the system (\ref{scalar} - \ref{gauge3}). We are presented
with a scenario where four different solutions exist:
\newline{\bf Normal phase}: an analytic solution where $\psi=w=0$,  $\Phi=\mu(1-1/r)$
and $\Theta=\mu_3(1-1/r)$ exists for any value of $\mu$ and $\mu_3$, and it describes the normal state of the
system.
\newline{\bf $s$-wave phase}: for $\mu-\mu_3\geq  8.127$ we find solutions with non-zero $\psi$ resembling those in the balanced case.
\newline{\bf $p$-wave phase}: for $|\mu_3|/\mu\geq3.65 /\mu$ solutions with $\psi=0$, but $w\neq 0$  satisfying
(\ref{uvconds2}) exist. The scalar condensate $\langle O_2\rangle$ is null while $\langle J_x^{(1)}\rangle\neq 0$.
These solutions break the $U(1)\times U(1) $ down to $U(1)$ and also break the $SO(2)$ corresponding to spatial
rotations. Notice that $w(r)$ is not charged under the overall $U(1)$ and therefore this solution is insensitive
to the value of $\mu$. This would change if the backreaction of the matter fields on the geometry was taken into account
as in \cite{Bigazzi:2011ak, Erdmenger:2011hp}.
\newline{\bf $s$+$p$-wave phase}: for small values of $\mu_3/\mu$ we find the extension of the $s$+$p$-wave
solution found in the previous section for $\mu_3=0$. However, the larger $|\mu_3|/\mu$ the larger the $\mu$
at which the phase appears. We have also found solutions with two condensates in an intermediate region in which
$\mu_3$ is large and $\mu$ is close to the critical value $\mu_s$. But they are always energetically
unfavored with respect to the pure $s$-wave solutions (see Figure \ref{phasediagram}).

\begin{figure}[htp]
\begin{center}
\includegraphics[width=3.4in]{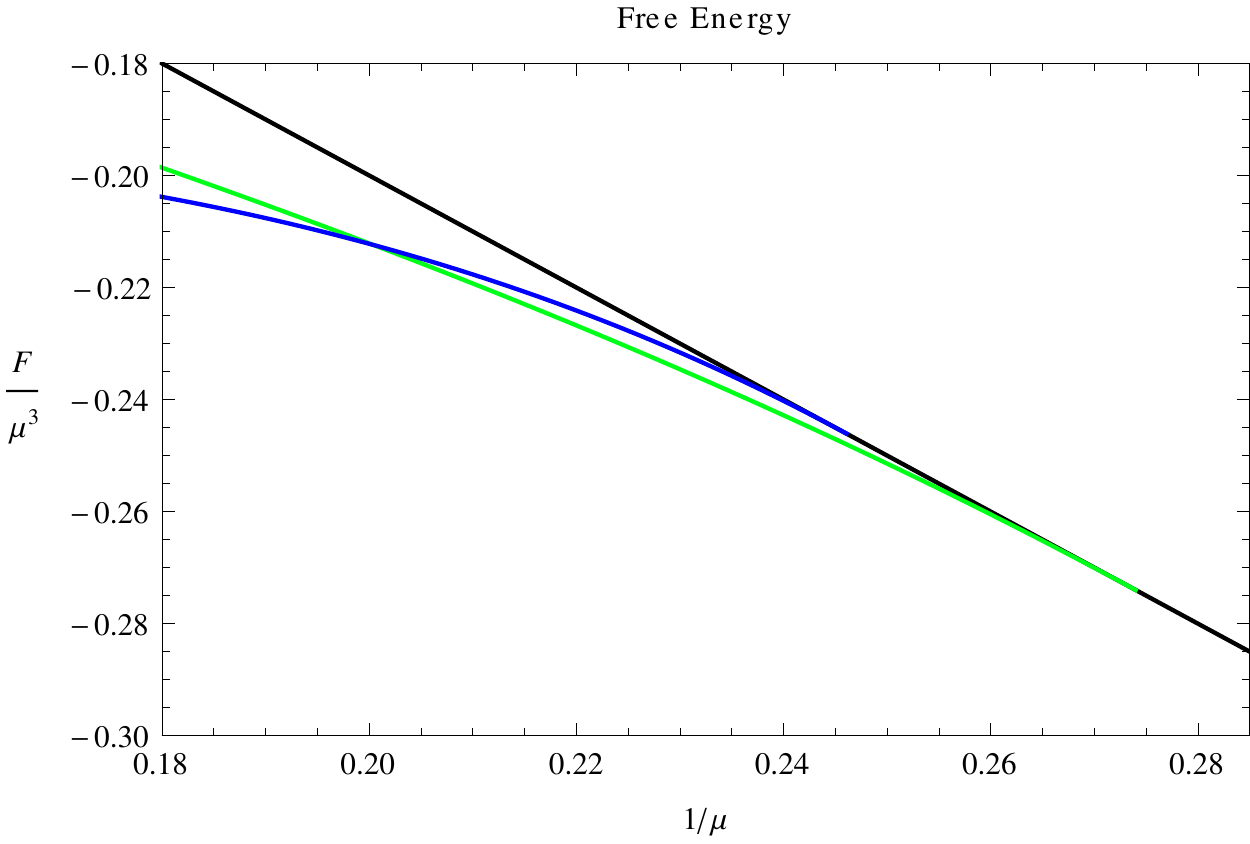}
\caption{\label{benchmark} Free energy as a function of $1/\mu$ for $\mu_3/\mu=-1$. Black corresponds to the normal phase,  blue to the $s$-wave phase, and green to the $p$-wave phase.}
\end{center}
\end{figure}

By computing the free energy (\ref{freeen}) of the different solutions we determine the phase diagram of the
system as a function of $1/\mu$ and $\mu_3/\mu$ which we plotted in figure \ref{phasediagram}. For small
values of $\mu_3/\mu$ the situation is very similar to what we found in the previous section for $\mu_3=0$.
As already mentioned, as $|\mu_3|/\mu$ gets larger, the transition to the $s$+$p$-wave phase happens at a
higher value of $\mu$. It might be the case that the phase eventually disappears at a finite value of that ratio, but
this would happen beyond the region of applicability of the decoupling limit, and thus backreaction should be taken
into account\footnote{Notice that if the $s$+$p$-wave phase survived down to $1/\mu =0$ for $\mu_3/\mu$ lower
than a critical value (as the phase diagram \ref{phasediagram} seems to imply) we would be in the pressence
of a quantum critical point at which the system goes from the $s$+$p$ to the $s$-wave phase. This resembles
what happens in \cite{sip} for the $p$+$is$ superconductor.}. For $|\mu_3|/\mu$ large enough, the $p$-wave phase is preferred at intermediate values of $\mu$. Therefore, as $\mu$ is increased above a critical value $\mu_p$ the system goes from the normal to the $p$-wave phase through a second order phase transition. If $\mu$ is increased even further a first order phase transition takes the system from the $p$-wave to the $s$-wave phase. This $p$- to $s$-wave first order phase transition is illustrated by figure \ref{benchmark} where we plot the free energy of both phases (and that of the normal phase) as a function of $\mu$ at a fixed
value of $\mu_3/\mu = -1$. The tricritical point where the normal, $s$-wave and $p$-wave phases meet happens at $1/\mu \approx 0.223$ and $|\mu_3|/\mu\approx 0.815$. The $p$-wave solution is never energetically preferred for $|\mu_3|/\mu< 0.815$.

A cautionary comment about the phase diagram of figure \ref{phasediagram} is in order. In the regions of the
parameter space where $|\mu_3|/\mu \gg1$ or $1/\mu\ll1$ the probe limit is not valid anymore, and therefore the
phase diagram might be modified once backreaction is taken into account \footnote{Remember that the decoupling
limit corresponds to taking the gauge coupling (and charge of the scalar field) $g_{\rm YM}$ to be very large,
so the effect of the matter fields on the metric is negligible. Hence it is valid as far as $\mu_i \ll  g_{\rm YM}$
and the condensates are small.}. Indeed, the nature of the different phase transitions, as well as the critical values of the chemical potentials could be altered in those regions \cite{Ammon:2009xh,Arias:2012py}. However, in $2+1$-dimensions both the $s$-wave and $p$-wave superconducting phase transitions separately are known to remain second order even lor large backreaction \cite{Bigazzi:2011ak, Erdmenger:2011hp}. Therefore, we expect the main features of the phase diagram like the existence of distinct $s$ and $p$-wave phases meeting at a tricritical point will not be very sensitive to backreaction. The order of the phase transition between the $s$ and $p$-wave phases could still be modified by backreaction.

\section{Conclusions}
In this paper we report on the construction of  a holographic $s$+$p$-wave superconducting state. This phase,
where both an $s$-wave and $p$-wave condensates exist, is the preferred state at low temperatures of the holographic
two-component superfluid first presented in \cite{we}. This model realizes a global $U(2)$ symmetry on the boundary
theory and presents superconducting states with non-vanishing charge density corresponding to the two different
$U(1)$s inside the $U(2)$.

Our main results are summarized by figures \ref{condensate} and \ref{phasediagram}. Figure \ref{condensate} shows
that an $s$+$p$-wave state appears at low temperatures. A free energy analysis determined that the system enters
this state through a second order phase transition, and stays in it for as low temperature as we can go.
On the other hand, figure \ref{phasediagram} presents the phase diagram for  the unbalanced system: chemical
potentials for the two $U(1)$s $\subset U(2)$ are turned on, and hence $U(2)$ is explicitly broken to
$U(1)\times U(1)$. In this phase diagram three different superconducting phases are present. These are the
standard $s$-wave phase where  a scalar condensate breaks the $U(1)\times U(1)$ down to $U(1)$; a $p$-wave phase
where $\langle J_x^{(1)}\rangle\neq0\,$, $U(1)\times U(1)$ is broken to (a different) $U(1)$, and also spatial
rotational symmetry is broken; and an $s$+$p$-wave phase where the $U(1)\times U(1)$ is completely broken by the
$s$ and $p$-wave condensates, and again spatial rotational symmetry is  broken. Remarkably, while the system goes
from the normal phase to the $s$-wave and $p$-wave phases through second order phase transitions, the phase
transition between the $s$ and $p$-wave phases is always a first order one. The existence of this first order phase
transition between superconducting phases in the unbalanced system is an interesting prediction of our holographic
model. These conclusions could be sensitive to the inclusion of backreaction since, as already mentioned,
in principle the order of the phase transitions could change when the parameters are large and the decoupling
limit breaks down. Yet in the proximity of the tricritical point, where the $p$- and $s$-wave phases meet,
the matter fields and its derivatives are small enough for the probe limit to be trusted. Hence the existence of
this point and the first order phase transition between the $p$- and $s$-wave phases in its proximity will survive
once backreaction is considered, at least for large enough gauge coupling. Moreover, a preliminary study of
backreacted solutions in that region supports this conclusion and show it holds for small values of the gauge
coupling too \cite{prep}.
In any case, in order to ensure the stability of the different phases it is important to study the quasinormal
mode spectrum of the model. As pointed out in \cite{us}, it might be possible that instabilities towards
inhomogeneous phases appear.

In \cite{Gusynin:2003yu} a QFT model featuring a gauged $U(2)$
symmetry, and with a
 symmetry breaking scheme similar to ours is studied. There the autors find roton excitations along the
 direction of the vector condensate.  It would be interesting to study the quasinormal mode spectra of the
 $s$+$p$-wave phase and see if something similar happens in our case. We leave this for a future investigation.


{\bf Acknowledgments}
We specially thank Karl Landsteiner for contributions at an early stage of this project and for valuable comments and discussions. We are grateful to Francesco Bigazzi, Daniele Musso, Leopoldo Pando Zayas, Antonello Scardicchio, and Paul Torr\'io for useful comments and discussions.  D.A. thanks the FRoGS for unconditional support. This work has been supported by MEC and FEDER grant FPA2012-32828, Consolider Ingenio Programme CPAN (CSD2007-00042), Comunidad de Madrid HEP-HACOS
S2009/ESP-1473 and MINECO Centro de excelencia Severo Ochoa Program under grant SEV-2012-0249. I. A. is supported by the Israel Science Foundation under grants no. 392/09 and 495/11. L.M. has been supported by FPI-fellowship BES-2010-041571. A. J. is supported by FPU fellowship AP2010-5686.


\begin{thebibliography}{100}

\bibitem{zhang}
  S.~C.~Zhang,
 Science, 1997:
Vol. 275 no. 5303 pp. 1089-1096


\bibitem{he3phases}
D. Vollhardt and P. Wolfle,
(Taylor \& Francis, London, 1990).

\bibitem{sip}
  P.~Goswami and B.~Roy,
  arXiv:1307.3240 [cond-mat.supr-con].

  \bibitem{ludwig}
A. P. Schnyder, S. Ryu, A. Furusaki, A. W. W. Ludwig,
Phys. Rev. B {\bf 78}, 195125 (2008).

\bibitem{Gubser:2008zu}
  S.~S.~Gubser,
  Phys.\ Rev.\ Lett.\  {\bf 101} (2008) 191601
  [arXiv:0803.3483 [hep-th]].

\bibitem{Hartnoll:2008vx}
  S.~A.~Hartnoll, C.~P.~Herzog and G.~T.~Horowitz,
  Phys.\ Rev.\ Lett.\  {\bf 101} (2008) 031601
  [arXiv:0803.3295 [hep-th]].

\bibitem{Hartnoll:2009sz}
  S.~A.~Hartnoll,
  Class.\ Quant.\ Grav.\  {\bf 26} (2009) 224002
  [arXiv:0903.3246 [hep-th]].

\bibitem{Horowitz:2010gk}
  G.~T.~Horowitz,
  Lect.\ Notes Phys.\  {\bf 828} (2011) 313
  [arXiv:1002.1722 [hep-th]].


\bibitem{Hartnoll:2008kx}
  S.~A.~Hartnoll, C.~P.~Herzog and G.~T.~Horowitz,
  JHEP {\bf 0812} (2008) 015
  [arXiv:0810.1563 [hep-th]].


\bibitem{phsc}
  S.~S.~Gubser and S.~S.~Pufu,
  JHEP {\bf 0811} (2008) 033
  [arXiv:0805.2960 [hep-th]];
  M.~Ammon, J.~Erdmenger, M.~Kaminski and P.~Kerner,
  Phys.\ Lett.\ B {\bf 680} (2009) 516
  [arXiv:0810.2316 [hep-th]].


\bibitem{dhsc}
   J.~-W.~Chen, Y.~-J.~Kao, D.~Maity, W.~-Y.~Wen and C.~-P.~Yeh,
  Phys.\ Rev.\ D {\bf 81} (2010) 106008
  [arXiv:1003.2991 [hep-th]];
  F.~Benini, C.~P.~Herzog, R.~Rahman and A.~Yarom,
  JHEP {\bf 1011} (2010) 137
  [arXiv:1007.1981 [hep-th]].

\bibitem{Basu:2010fa}
  P.~Basu, J.~He, A.~Mukherjee, M.~Rozali and H.~-H.~Shieh,
  JHEP {\bf 1010} (2010) 092
  [arXiv:1007.3480 [hep-th]].


  \bibitem{Zayas:2011dw}
  L.~A.~Pando Zayas and D.~Reichmann,
  Phys.\ Rev.\ D {\bf 85} (2012) 106012
  [arXiv:1108.4022 [hep-th]].

\bibitem{Musso:2013ija}
  D.~Musso,
  JHEP {\bf 1306} (2013) 083
  [arXiv:1302.7205 [hep-th]].

\bibitem{Cai:2013wma}
  R.~-G.~Cai, L.~Li, L.~-F.~Li and Y.~-Q.~Wang,
  arXiv:1307.2768 [hep-th].

\bibitem{Wen:2013ufa}
  W.~-Y.~Wen, M.~-S.~Wu and S.~-Y.~Wu,
  arXiv:1309.0488 [hep-th].

\bibitem{Nie:2013sda}
  Z.~-Y.~Nie, R.~-G.~Cai, X.~Gao and H.~Zeng,
  arXiv:1309.2204 [hep-th].

\bibitem{Amoretti:2013oia}
  A.~Amoretti, A.~Braggio, N.~Maggiore, N.~Magnoli and D.~Musso,
  arXiv:1309.5093 [hep-th].
  
    \bibitem{Donos:2013woa}
   A.~Donos, J.P.~Gauntlett and C.~Pantelidou,
   arXiv:1310.5741 [hep-th].

\bibitem{we}
  I.~Amado, D.~Arean, A.~Jimenez-Alba, K.~Landsteiner, L.~Melgar and I.~S.~Landea,
  JHEP {\bf 1307} (2013) 108
  [arXiv:1302.5641 [hep-th]].

\bibitem{halperin}
B. Halperin,
Phys. Rev. B {\bf 11}, 178–190 (1975).

\bibitem{imfermi}
Y.-i. Shin, C. H. Schunck, A. Schirotzek, and W. Ketterle,
Nature 451 (2008) 689693.

\bibitem{imqcd}
L.~-y.~He, M.~Jin and P.~-f.~Zhuang,
  Phys.\ Rev.\ D {\bf 71} (2005) 116001
  [hep-ph/0503272];
   M.~N.~Chernodub and A.~S.~Nedelin,
  Phys.\ Rev.\ D {\bf 83} (2011) 105008
  [arXiv:1102.0188 [hep-ph]].


\bibitem{loffrv}
R. Combescot,
arXiv:cond-mat/0702399v1 [cond-mat.supr-con].

\bibitem{loff}
P. Fulde and R. A. Ferrell,
Phys. Rev. 135 (1964) A550;
A. I. Larkin and Y. N. Ovchinnikov,
ZhEFT, 47 (1964) 1136 [Sov. Phys. JETP, 20 (1965) 762].

\bibitem{Bigazzi:2011ak}
  F.~Bigazzi, A.~L.~Cotrone, D.~Musso, N.~P.~Fokeeva and D.~Seminara,
  JHEP {\bf 1202} (2012) 078
  [arXiv:1111.6601 [hep-th]].


\bibitem{Erdmenger:2011hp}
  J.~Erdmenger, V.~Grass, P.~Kerner and T.~H.~Ngo,
  JHEP {\bf 1108} (2011) 037
  [arXiv:1103.4145 [hep-th]].


\bibitem{Krikun:2012yj}
  A.~Krikun, V.~P.~Kirilin and A.~V.~Sadofyev,
  JHEP {\bf 1307} (2013) 136
  [arXiv:1210.6074 [hep-th]].

\bibitem{prep}
I.~Amado, D.~Arean, A.~Jimenez-Alba, L.~Melgar and I.~S.~Landea,
{\it In preparation}.
  
  
\bibitem{us}
  I.~Amado, D.~Arean, A.~Jimenez-Alba, K.~Landsteiner, L.~Melgar and I.~S.~Landea,
  arXiv:1307.8100 [hep-th].

\bibitem{Ammon:2009xh}
  M.~Ammon, J.~Erdmenger, V.~Grass, P.~Kerner and A.~O'Bannon,
  Phys.\ Lett.\ B {\bf 686} (2010) 192
  [arXiv:0912.3515 [hep-th]].

\bibitem{Arias:2012py}
  R.~E.~Arias and I.~S.~Landea,
  JHEP {\bf 1301} (2013) 157
  [arXiv:1210.6823 [hep-th]].


\bibitem{Gusynin:2003yu}
  V.~P.~Gusynin, V.~A.~Miransky and I.~A.~Shovkovy,
  Phys.\ Lett.\ B {\bf 581} (2004) 82
  [hep-ph/0311025].

\end{thebibliography}
\end{document}